\documentclass[structabstract]{aa}
\usepackage{txfonts}
\usepackage{graphicx}
\usepackage{natbib}
\bibpunct{(}{)}{;}{a}{}{,} % to follow the A&A style
\begin{document}
\title{Distinctive rings in the 21 cm signal of the epoch of reionization}

\author{P. Vonlanthen \inst{1} \and B. Semelin \inst{1,2} \and S. Baek \inst{3} \and Y. Revaz \inst{4}}

\institute{LERMA, Observatoire de Paris, 61 Av. de l'Observatoire, 75014 Paris, France\\ \email{patrick.vonlanthen@obspm.fr} \and Universit\'e Pierre et Marie Curie, 4 Place Jules Janssen, 92195 Meudon Cedex, France \and Scuola Normale Superiore, Piazza dei Cavalieri 7, 56126 Pisa, Italy \and Laboratoire d'Astrophysique, Ecole Polytechnique F\'ed\'erale de Lausanne (EPFL), Switzerland
}

\date{Received / Accepted}

\abstract
%Context
{It is predicted that sources emitting UV radiation in the Lyman band during the epoch of reionization showed a series of discontinuities in their Ly$\alpha$ flux radial profile as a consequence of the thickness of the Lyman line series in the primeval intergalactic medium. Through unsaturated Wouthuysen-Field coupling, these spherical discontinuities are also present in the 21 cm emission of the neutral IGM.%If they are shown to be detectable with the planned Square Kilometre Array, they would provide an unambiguous diagnostic for the cosmological origin of the signal.
}
%Aims
{In this article, we study the effects these discontinuities have on the differential brightness temperature of the 21 cm signal of neutral hydrogen in a realistic setting including all other sources of fluctuations. We focus on the early phases of the epoch of reionization, and we address the question of the detectability by the planned Square Kilometre Array. Such a detection would be of great interest, because these structures could provide an unambiguous diagnostic for the cosmological origin of the signal remaining after the foreground cleaning procedure. %diagnostic to disentangle signals of cosmological origin from residual from imperfect foreground removal.
Also, they could be used as a new type of standard rulers.}
%Methods
{We determine the differential brightness temperature of the 21 cm signal in the presence of inhomogeneous Wouthuysen-Field effect using simulations which include (hydro)dynamics and both ionizing and Lyman lines 3D radiative transfer with the code LICORICE. We include radiative transfer for the higher-order Lyman-series lines and consider also the effect of backreaction from recoils and spin diffusivity on the Ly$\alpha$ resonance.}
%Results
{We find that the Lyman horizons are difficult to indentify using the power spectrum of the 21 cm signal but are clearly visible on the maps and radial profiles around the first sources of our simulations, but for a limited time interval, typically $\Delta z \approx 2$ at $z \sim 13$. Stacking the profiles of the different sources of the simulation at a given redshift results in extending this interval to $\Delta z \approx 4$. When we take into account the implementation and design planned for the SKA (collecting area, sensitivity, resolution), we find that detection will be challenging. It may be possible with a 10 km diameter for the core, but will be difficult with the currently favored design of a 5 km core. %While instrumental noise can strongly affect the horizon detectability around individual sources at the highest redshifts, we find that, as soon as the number of sources is sufficient, the stacking procedure is still efficient.
}
%Conclusions
{%Detection of differential brightness temperature discontinuities on 21 cm maps should be possible with the SKA if the first sources are mainly stars. Observational attempts would be worth time and effort involved because these structures could provide a diagnostic to disentangle signals of cosmological origin from residual from imperfect foreground removal. Also, they could be used as a new type of standard rulers.
}


\maketitle
\newcommand{\Lya}{Ly$\alpha$ }
\newcommand{\Lyb}{Ly$\beta$ }
\newcommand{\Lyg}{Ly$\gamma$ }
\newcommand{\Lyd}{Ly$\delta$ }
\newcommand{\Lye}{Ly$\epsilon$ }
\newcommand{\Lyz}{Ly$\zeta$ }
\newcommand{\Lyn}{Ly$n$ }
\newcommand{\WF}{Wouthuysen-Field }
\newcommand{\tk}{$T_{\rm{k}}$}
\newcommand{\tr}{$T_{\gamma}$}
\newcommand{\tsp}{$T_{\rm{s}}$}
\newcommand{\dtb}{$\delta T_{\mathrm{b}}$ }

%---------------------------------------------------------------------------------------
\section{Introduction}
%---------------------------------------------------------------------------------------
Our universe was initially fully ionized, from the primordial very hot and dense phase until cosmological recombination, occurring at $z \sim 1100$. After recombination, baryonic matter is made of a neutral gas of hydrogen and helium . This period, known as the Dark Ages, lasted until the first light sources lit up at lower redshifts, as a consequence of the growth of primordial density fluctuations. These sources are responsible for the subsequent reionization of the cosmological gas. Numerous questions still have to be answered on that Epoch of Reionization (EoR), and to date only a few observational results are in a position to put constraints on that period. The Gunn-Peterson effect \citep{GunnPeterson65} indicates that the neutral hydrogen fraction is lower that $10^{-4}$ at $z < 5.5$ \citep{fanetal06}\footnote{For an interesting discussion about the redshift at which reionization is complete, as deduced from the spectra of high redshift quasars, see \citet{Mesinger10}.}. Another constraint is the optical depth due to Thomson scattering of cosmic microwave background (CMB) photons by free electrons. The seven-year results of the Wilkinson Microwave Anisotropy Probe (WMAP) experiment, combined with baryon acoustic oscillation data from the Sloan Digital Sky Survey and priors on $H_0$ from Hubble Space Telescope observations, yield an optical depth $\tau = 0.088 \pm 0.014$, which implies a reionization redshift of $z = 10.6 \pm 1.2$ \citep{Komatsuetal11}, assuming an instantaneous transition. From these two observational results, we can infer that the EoR lasted most likely over an extended period. In addition to these two constraints, new observations of the UV luminosity functions of galaxies in the reionization epoch have recently been published. These studies are in a position to put constraints on the efficiency of galaxies in reionizing the intergalactic medium (IGM). See for example the recent HUDF results at $z \approx 7-10$ \citep{Bouwensetal11,Bouwensetal10}.

The most promising probe of the EoR, however, is the future observation of the 21 cm hyperfine line of neutral hydrogen (H\textsc{i}). Indeed, H\textsc{i} can be detected in emission or absorption against the CMB at the wavelength of the redshifted 21 cm line corresponding to the transition between the two hyperfine levels of the electronic ground state \citep{HoganRees79,ScottRees90}. Pathfinder experiments such as LOFAR\footnote{\texttt{http://www.lofar.org/}}, MWA\footnote{\texttt{http://www.haystack.mit.edu/ast/arrays/mwa/}}, GMRT\footnote{\texttt{http://gmrt.ncra.tifr.res.in/}}, 21CMA\footnote{\texttt{http://21cma.bao.ac.cn/}} or PAPER\footnote{\texttt{http://astro.berkeley.edu/$\sim$dbacker/eor/}} will deduce information about the statistical properties of the signal, while the proposed Square Kilometre Array (SKA\footnote{\texttt{http://www.skatelescope.org/}}) will be able to deliver a full tomography of the IGM.

An accurate prediction of the 21 cm signal requires the determination of the baryonic density, the ionization fraction, the kinetic temperature and the local \Lya flux. Numerous parameters can strongly affect the properties of the predicted signal, such as the size of the simulation box \citep{Ciardietal03b,BarkanaLoeb04,Ilievetal06,Baeketal09}, the effect of the \Lya wing scattering \citep{Semelinetal07,ChuzhoyZheng07}, or the nature of the reionizing sources \citep{Ciardietal03a,Mellemaetal06,Ilievetal07,VolonteriGnedin09,Baeketal10}. Depending on these parameters, the EoR can extend over a variable redshift range and produce a patchy morphology with sharper or smoother transitions between neutral and ionized periods \citep{Furlanettoetal06}. Knowledge of the \Lya flux is also important, because it allows one to compute precisely the spin temperature of neutral hydrogen. Indeed, observations of the redshifted 21 cm line are only possible when the spin temperature is different from the CMB temperature, and among the physical processes likely to decouple these two temperatures, the expected dominant one is the \WF effect \citep{Wouthuysen52,Field58}, i.e. the pumping of the 21 cm transition by \Lya photons. For that reason, several studies focused on the \Lya radiative transfer during the EoR \citep{BarkanaLoeb05,Furlanetto06,Semelinetal07,ChuzhoyZheng07,Baeketal09}.

In this paper we will consider the role of the radiative transfer of higher-order Lyman-series photons in the determination of the differential brightness temperature of the 21 cm signal. This is motivated by the fact that the local \Lya photon intensity at a given redshift is made of two contributions: photons that were emitted below \Lyb and redshifted to the local \Lya wavelength; and photons emitted between \Lyb and the Lyman limit, that redshifted until they reached the nearest atomic level $n$. Due to the fact that the IGM is optically thick also to the higher levels, these photons were quickly absorbed and reemitted. But during these repeated scatterings the excited electrons pretty soon cascaded toward lower levels, a fraction of these radiative cascades ending with the emission of a \Lya photon.

The contribution of higher-order Lyman-line photons has been previously studied by several authors.  However, while the first studies assumed a 100 percent efficiency for radiative cascades to end as a \Lya photon \citep[e.g.][]{BarkanaLoeb05}, \citet{Hirata06} and \citet{PritchardFurlanetto06} showed that a sizeable fraction of the cascading photons ends in a two-photon decay from the 2s level. These authors calculated the exact cascade conversion probabilities and discussed the possibility of hyperfine level mixing by scattering of \Lyn photons.

As a consequence of the thickness of the Lyman resonances in the primeval IGM, any primordial light source emitting photons in the Lyman band is surrounded by discrete horizons, i.e. maximum distances photons with a given emission frequency can travel before redshifting into a Lyman resonance and being scattered by neutral hydrogen. Thus, \Lya flux profiles around ionizing sources exhibit characteristic discontinuities at the Lyman horizons. The importance of the \WF effect in the determination of the differential brightness temperature \dtb of the 21 cm signal during the process of cosmic reionization motivates us to tackle the question of detecting corresponding brightness temperature discontinuities around the first light sources. In this paper, we include for the first time a correct treatment of the radiative transfer of higher-order Lyman-series lines into realistic numerical simulations of the EoR. The question is: will the horizons be observable in brightness temperature data over all the other sources of fluctuations, either physical (density, temperature, velocity effects) or instrumental? If this is the case, these spherical horizons of known radii at a given redshift will provide a diagnostic to test systematics and foreground removal procedures \citep[see e.g.][]{Harkeretal09}. Without such a test, residual from imperfect removal may be impossible to disentangle from the cosmological signal.

Our paper is organized as follows. In Sect. \ref{WFE} we give some generalities about the \WF effect and the differential brightness temperature of the 21 cm transition. Lyman horizons are discussed in Sect. \ref{Lyhor}. Details on our numerical simulations are described in Sect. \ref{sim}, while Sect. \ref{res} shows our results. Finally, we set out our conclusions in Sect. \ref{concl}.
%---------------------------------------------------------------------------------------

%---------------------------------------------------------------------------------------
\section{\WF effect and differential brightness temperature of the 21 cm transition}
\label{WFE}
%---------------------------------------------------------------------------------------
The differential brightness temperature of the 21 cm hyperfine transition is determined by the spin temperature $T_{\mathrm{s}}$, which is related to the ratio between the densities of hydrogen atoms in the triplet ($n_1$) and singlet ($n_0$) hyperfine levels of the electronic ground state:
\begin{equation}
\frac{n_1}{n_0} = \frac{g_1}{g_0} e^{-T_\star/T_{\mathrm{s}}},
\end{equation}
where $g_1/g_0 = 3$ is the ratio of the statistical weights and $T_\star = h\nu_{10}/k_\mathrm{B} = 0.0682$ K is the temperature corresponding to the energy difference between the two hyperfine levels, with $h$ the Planck constant, $\nu_{10} = 1420.4057$ MHz the hyperfine transition frequency and $k_\mathrm{B}$ the Boltzmann constant. Three processes can excite these levels: absorption of CMB photons, collisions, and the \WF effect, i.e. the mixing of the hyperfine levels through absorption and spontaneous reemission of \Lya photons. The latter effect dominates once the first luminous sources appear. Thus, the spin temperature can be written as \citep[see e.g.][]{Furlanettoetal06}:
\begin{equation}
T^{-1}_{\mathrm{s}} = \frac{T^{-1}_{\gamma}+x_{\alpha}(T_{\mathrm{c}}^{\mathrm{eff}})^{-1}+x_{\mathrm{c}}T^{-1}_{\mathrm{k}}}{1+x_{\alpha}+x_{\mathrm{c}}},
\label{equts}
\end{equation}
where $T_{\gamma}$ is the CMB temperature, $T_{\mathrm{c}}^{\mathrm{eff}}$ the effective color temperature of the UV radiation field, $T_{\mathrm{k}}$ the gas kinetic temperature, $x_{\alpha}$ the coupling coefficient for \Lya pumping, and $x_{\mathrm{c}}$ the coupling coefficient for collisions.

The coefficient $x_{\mathrm{c}}$ is given by:
\begin{equation}
x_{\mathrm{c}} = \frac{T_{\star}}{A_{10}T_{\gamma}}(C_{\mathrm{H}}+C_{\mathrm{p}}+C_{\mathrm{e}}),
\end{equation}
with $A_{10} = 2.85 \times 10^{-15}$ s$^{-1}$ the spontaneous emission factor of the 21 cm transition. $C_{\mathrm{H}}$, $C_{\mathrm{p}}$ and $C_{\mathrm{e}}$ are the de-excitation rates due to collisions with neutral hydrogen atoms, protons and electrons respectively, for which we use the fitting formulae given by \citet{Kuhlenetal06}.

The coupling coefficient for \Lya pumping, $x_\alpha$, is given by:
\begin{equation}
x_\alpha = \frac{4 P_\alpha T_\star}{27 A_{10} T_{\gamma}},
\end{equation}
where $P_\alpha$ is the \Lya scattering rate per atom per second \citep{Madauetal97}. In relation to the local \Lya flux, the coupling coefficient can also be written as:
\begin{equation}
x_\alpha = \frac{16 \pi^2 T_\star e^2 f_\alpha}{27 A_{10} T_{\gamma} m_\mathrm{e} c} S_\alpha J_\alpha,
\end{equation}
with $e$ the electron charge, $f_\alpha$ the \Lya oscillator strength, $m_\mathrm{e}$ the electronic mass, $c$ the speed of light and $J_\alpha$ the angle-averaged specific intensity of \Lya photons by photon number. The backreaction factor $S_\alpha$ accounts for spectral distortions near the \Lya resonance, sourced by recoils and spin diffusivity. This factor cannot be neglected at the low kinetic temperatures characterizing the very first stages of the EoR, when heating of the IGM by the first sources is still negligible. We include this important correction factor using the simple analytical fit from Hirata (2006), accurate to within 1\% in the range $T_{\mathrm{k}} \ge 2$K, $T_{\mathrm{s}} \ge 2$K, and $10^5 < \tau_{\rm{GP}} < 10^7$, $\tau_{\rm{GP}}$ being the Gunn-Peterson optical depth in \Lya \citep{GunnPeterson65}.

When photons emitted with frequencies between \Lyb and the Lyman limit redshift to the nearest atomic level $n$, they soon induce a radiative cascade, because of the thickness of the IGM to all Lyman levels. A fraction of these cascades ends with the emission of a \Lya photon. \citet{Hirata06} and \citet{PritchardFurlanetto06} showed that this process is not very efficient in producing \Lya photons, most of the radiative cascades ending in the 2s level, from which reaching the ground level is only possible through a two-photon decay. Furthermore, excitations of the 2s level via radiative excitations by CMB photons or collisional excitations are negligible at $z < 400$. As a consequence, these photons are lost for the \WF effect. Besides, it should also be noted that \textit{direct} contribution of \Lyn scattering to the coupling of the spin temperature to the gas temperature is not significant, as was shown by \citet{PritchardFurlanetto06}, the number of scattering events before cascading being about 5, much less than the number of \Lya scatterings through the line.

Once the spin temperature is known, the differential brightness temperature of the 21 cm signal can be determined by the following expression \citep[see e.g.][]{Madauetal97,CiardiMadau03}:
\begin{eqnarray}
\nonumber \delta T_{\mathrm{b}} & = & 28.1 \ x_{\mathrm{HI}} \ (1 + \delta) \left( \frac{1+z}{10} \right)^{1/2} \frac{T_{\mathrm{s}}-T_{\gamma}}{T_{\mathrm{s}}}\\
&& \times \left( \frac{\Omega_{\mathrm{b}}}{0.042} \frac{h}{0.73} \right) \left( \frac{0.24}{\Omega_{\mathrm{m}}} \right)^{1/2} \left( \frac{1 - Y_{\mathrm{p}}}{1-0.248} \right) \mathrm{mK},
\label{equdtb}
\end{eqnarray}
where $x_{\mathrm{HI}}$ is the neutral hydrogen fraction, ($1 + \delta$) the fractional baryon overdensity at redshift $z$, and $\Omega_{\mathrm{b}}$, $\Omega_{\mathrm{m}}$, $h$, and $Y_{\mathrm{p}}$ the usual cosmological parameters. Note that we do not consider in this paper any contribution from the proper velocity gradient.
%---------------------------------------------------------------------------------------

%---------------------------------------------------------------------------------------
\section{Lyman horizons}
\label{Lyhor}
\begin{figure}[!t]
\resizebox{\hsize}{!}{\includegraphics[angle=270]{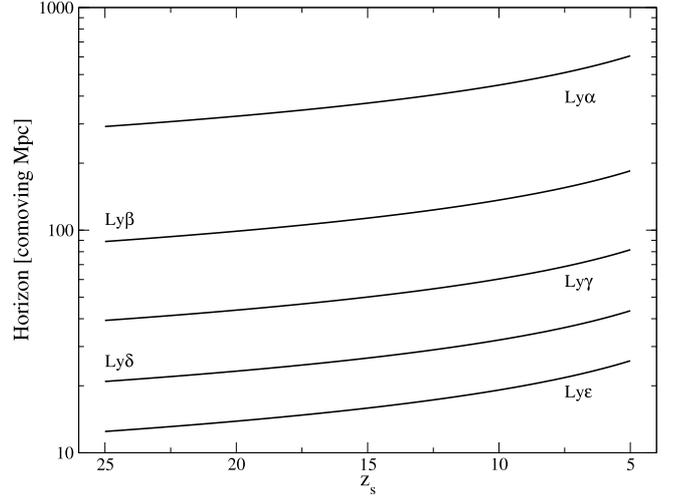}}
\caption{Evolution of the first five Lyman horizons, in comoving Mpc, as a function of the emission redshift. The horizon labeled \Lyn is the maximal distance a photon emitted just below Ly$(n+1)$ can travel.}
\label{fighor}
\end{figure}

%---------------------------------------------------------------------------------------
Photons emitted by a source at redshift $z_{\mathrm{s}}$ with frequency $\nu_{\mathrm{s}}$ are observed with frequency $\nu_{\mathrm{obs}}$ at $z_{\mathrm{obs}}$, with the following relation between emitted and observed frequencies:
\begin{equation}
\frac{\nu_{\mathrm{s}}}{\nu_{\mathrm{obs}}} = \frac{1+z_{\mathrm{s}}}{1+z_{\mathrm{obs}}}.
\end{equation}
The comoving separation $d$ between the source and the observer is determined by:
\begin{equation}
d = c \int\limits_{t_{\mathrm{s}}}^{t_{\mathrm{obs}}} \frac{1}{a(t)} \mathrm{d}t = c \int\limits_{z_{\mathrm{obs}}}^{z_{\mathrm{s}}} \frac{1}{H(z)} \mathrm{d}z,
\label{hor}
\end{equation}
with $a$ the scale factor and $H$ the Hubble parameter. At sufficiently high redshifts, the contribution of the cosmological constant to $H$ can safely be neglected, and we can approximate $H(z) \approx H_0 (1+z) \sqrt{\Omega_{\mathrm{m}} (1+z)}$, with $H_0$ the present-day value for the Hubble parameter. In that case, equation (\ref{hor}) leads to:
\begin{equation}
d = \frac{2c}{H_0 \sqrt{\Omega_{\mathrm{m}}}} \left( 1 + z_{\mathrm{s}} \right)^{-1/2} \left( \left( \frac{\nu_{\mathrm{obs}}}{\nu_{\mathrm{s}}}  \right)^{-1/2} - 1 \right).
\end{equation}
The maximum distance a photon emitted just below Ly$(n+1)$ can travel before reaching the \Lyn resonance is obtained when using $\nu_{\mathrm{s}} = \nu_{\mathrm{n+1}}$ and $\nu_{\mathrm{obs}} = \nu_{\mathrm{n}}$, with $\nu_n$ the frequency corresponding to the transition from level $n$ to the ground state:
\begin{equation}
\nu_n = \nu_{\mathrm{LL}} (1 - n^{-2}),
\end{equation}
where $\nu_{\mathrm{LL}}$ is the Lyman limit frequency. We then have the following expression for the \Lyn horizons:
\begin{equation}
d_n = \frac{2c}{H_0 \sqrt{\Omega_{\mathrm{m}}}} \left( 1 + z_{\mathrm{s}} \right)^{-1/2} \left( \left( \frac{\nu_n}{\nu_{n+1}} \right)^{-1/2} - 1 \right), \qquad n \ge 2.
\label{hor2}
\end{equation}
%Note, however, that the change in the Hubble parameter during the travel time of the photons is small. For that reason, $H$ can be considered as constant in equation (\ref{hor}), and equation (\ref{hor2}) reduces to:
%\begin{equation}
%d_n \approx \frac{c}{H(z_{\mathrm{s}})} \left( 1 + z_{\mathrm{s}} \right) \left( 1 - \frac{\nu_{n}}{\nu_{n+1}} \right), \qquad n \ge 2.
%\label{horapprox}
%\end{equation}
Figure \ref{fighor} shows the evolution of the first five Lyman horizons, in comoving Mpc, as a function of the emission redshift. We see that photons emitted just below \Lyb can travel more than 350 comoving Mpc at $z \sim 15$, while photons emitted just below \Lyz cannot go farther than about 16 comoving Mpc. %The approximate values (dashed lines) given by equation (\ref{horapprox}) are close to the exact solutions (solid lines) of equation (\ref{hor2}), except for the \Lya horizon. The agreement between the two calculations increases with the level $n$ of the Lyman resonance.

The existence of these Lyman horizons around primordial sources during the early EoR (while the Wouthuysen-Field coupling is not yet saturated) leads to discontinuities in their \Lya flux profiles, which could create similar features in the brightness temperature profiles and maps. In order to answer quantitatively this question, we have to consider a full modelling of the \Lya pumping in a realistic numerical simulation of the EoR.
%---------------------------------------------------------------------------------------

%---------------------------------------------------------------------------------------
\section{Numerical methods}
\label{sim}
%---------------------------------------------------------------------------------------
We simulate cosmological reionization as a three step process. First, a dynamical simulation is run using GADGET2 \citep{Springel05}. Snapshots of the simulation are produced at regular intervals. The interval is given in terms of the scale factor: $\Delta a = 2^{-10}$. Then the Monte-Carlo 3D radiative transfer code LICORICE \citep{Semelinetal07,Baeketal09,Ilievetal09} is used to compute the ionizing continuum radiative transfer. We use two different dynamical simulations. The first one simulates a $( 100 \ h^{-1}$ Mpc$)^3$ volume with $2\times 256^3$ particles and assumes WMAP3 data alone cosmological parameters: $\Omega_{\mathrm{m}} = 0.24$, $\Omega_{\Lambda} = 0.76$, $\Omega_{\mathrm{b}} = 0.042$, $h = 0.73$ \citep{Spergeletal07}. The source modelling (stellar sources, Salpeter IMF, with masses in the range 1-120 M$_{\odot}$) and detailed reionization scheme, is described in \citet{Baeketal10}, where detailed analysis of the reionization history can be found. The second simulation is a box of size $200 \ h^{-1}$ Mpc, with $2\times 512^3$ particles, such as the mass resolution is the same as in the first one. It assumes the more recent WMAP7+BAO+$H_0$ cosmological parameters: $\Omega_{\mathrm{m}} = 0.272$, $\Omega_{\Lambda} = 0.728$, $\Omega_{\mathrm{b}} = 0.0455$, $h = 0.704$ \citep{Komatsuetal11}. Finally, the Lyman-line radiative transfer part of LICORICE is used as a second post-process to study the \WF effect, determine the spin temperature, and calculate the differential brightness temperature of the 21 cm signal of neutral hydrogen. In order to do so, we interpolated the output of the two reionization runs on a 256$^3$ (512$^3$) grid for the $100 \ h^{-1}$ ($200 \ h^{-1}$) Mpc simulation respectively. In both cases, we emitted $1.6 \times 10^9$ Lyman band photons between each pair of snapshots. We emphasize that the full radiative transfer in the lines is computed up to Ly$\epsilon$. Using an isotropic $r^{-2}$ profile around the sources would be much easier and faster. However, it has been shown that using the real, local \Lya flux is important at early redshifts, for ionization fractions smaller than $\sim 10$\%, when the \Lya coupling is not yet full \citep{ChuzhoyZheng07,Baeketal09}.

The implementation of radiative transfer for the higher-order Lyman-series lines is similar to the \Lya radiative transfer described in \citet{Semelinetal07} and \citet{Baeketal09}. Important modifications are listed here:
\begin{enumerate}
\item We now consider photons with frequencies in the range $[\nu_\alpha,\nu_\zeta]$, in order to include radiative transfer of the first five Lyman resonances (\Lya - Ly$\epsilon$). A 'flat' spectrum is assumed in this frequency range, i.e. every frequency has the same probability to be drawn\footnote{Given the narrowness of this frequency range, the use of a realistic non-flat spectrum results in minor differences, as we checked using blackbody spectra with three different effective temperatures: $5 \times 10^4$ K, $10^5$ K, and $1.5 \times 10^5$ K. \citet{ChuzhoyZheng07} also showed that the spectral slope of the source does not have a strong influence on the slope of the scattering rate profile.}. We emit $1.6 \times 10^9$ photons between each pair of snapshots.
\item When calculating the optical depth of the propagating photon, the \Lyn scattering cross-section is given by the Lorentzian profile:
\begin{equation}
\sigma_n(\nu) = f_{1n} \ \frac{\pi e^2}{m_{\mathrm{e}}c} \ \frac{\Delta \nu^n_{\mathrm{L}}/2 \pi}{\left( \nu - \nu^n_0 \right)^2 + \left( \Delta \nu^n_{\mathrm{L}} / 2 \right)^2},
\end{equation}
with $f_{1n}$ the \Lyn oscillator strength, $\nu^n_0$ the line center frequency, and $\Delta \nu^n_{\mathrm{L}}$ the natural line width. Note that for photons whose frequency lies between \Lyn and Ly$(n+1)$, we consider the possibility for scattering not only in the \Lyn line wing, but also in the Ly$(n+1)$ line wing. However, we find that scatterings in the Ly$(n+1)$ line wing are rare and can safely be neglected.
\item When a diffusion occurs in the \Lyn ($n > 2$) line, we consider the probability for cascading to Ly$\alpha$. The probability for a \Lyn photon to cascade is given by the value $p = 1 - P_{n\mathrm{p}\rightarrow 1s}$, where $P_{n\mathrm{p}\rightarrow 1s}$ is the simple diffusion probability from $n$ to the ground state \citep{PritchardFurlanetto06}. If the photon is not cascading, it is scattered by the atom.
\item For cascading photons, we take into account the recycling fraction $f_{\mathrm{recycle}}$ of radiative cascades that end in a \Lya photon emission. This fraction is zero for photons redshifting toward \Lyb and about 1/3 for photons redshifting toward Ly$n$, with $n \ge 4$ \citep{Hirata06,PritchardFurlanetto06}. The remaining radiative cascades do not contribute to the \WF effect, since they end in the 2s level, from which they can reach the ground state only with a forbidden 2$\gamma$-decay. We compute scatterings until photons cascade or redshift within 2 thermal linewidths of a \Lyn line center. At that distance from the center, given the thickness of the \Lyn lines, photons are assumed to scatter without spatial diffusion and initiate a radiative cascade on the spot.
\item Photons emitted below \Lyb which reach the \Lya resonance undergo $\tau_{\mathrm{GP}}$ diffusions until they redshift out of the line, with:
\begin{equation}
\tau_{\mathrm{GP}} = \frac{3 n_{\mathrm{H}} x_{\mathrm{HI}} \lambda^3_{\mathrm{Ly\alpha}} \gamma}{2H},
\end{equation}
where $n_{\mathrm{H}}$ is the proper number density of hydrogen, $\lambda_{\mathrm{Ly\alpha}}$ the \Lya wavelength, and $\gamma = 50$ MHz the half width at half-maximum of the \Lya resonance. Typical values are $\tau_{\mathrm{GP}} \sim 10^6$ at $z \sim 10$. On the other hand, cascading photons are injected in the center of the \Lya line\footnote{Indeed, \citet{FurlanettoPritchard06} noted that even if the first absorption is well blueward of the Lyman-series line center, the intermediate states of the atom during the radiative cascade have small natural widths.} and thus undergo $\frac{1}{2} \tau_{\mathrm{GP}}$ diffusions.
\end{enumerate}

In \citet{Baeketal09}, the photon propagation was stopped at 10 thermal linewidths of the \Lya line center. Indeed, at that distance from the center, the photon mean free path is much less than 1 physical kpc, assuming standard density and temperature conditions for the redshifts of interest. However, the mean free path is about 4 physical kpc at 10 thermal linewidths of the \Lye line center. We thus decided to stop the photon propagation at 2 thermal linewidths, where the mean free paths are much shorter: $\sim 10^{-5}$ and $\sim 7 \times 10^{-4}$ physical kpc for \Lya and \Lye photons respectively.

Note that we consider that \Lya heating is negligible \citep{PritchardFurlanetto06,FurlanettoPritchard06}. In a recent paper, \citet{Meiksin10} studies heating by higher-order Lyman-series photons and gets heating rates as high as several tens of degrees per Gyr. However, we do not take this process into account, as our work concentrates on the very early phases of the reionization epoch. For the same reason, we completely neglect the effect of dust. While dust may play a significant role in Lyman-line radiative transfer during the later stages of the EoR, this is not expected in the beginning of the EoR and we consider a fully dust-free universe.
%---------------------------------------------------------------------------------------

%---------------------------------------------------------------------------------------
\section{Results}
\label{res}
%---------------------------------------------------------------------------------------
\subsection{3D power spectra}
\begin{figure}[!t]
\resizebox{\hsize}{!}{\includegraphics[angle=270]{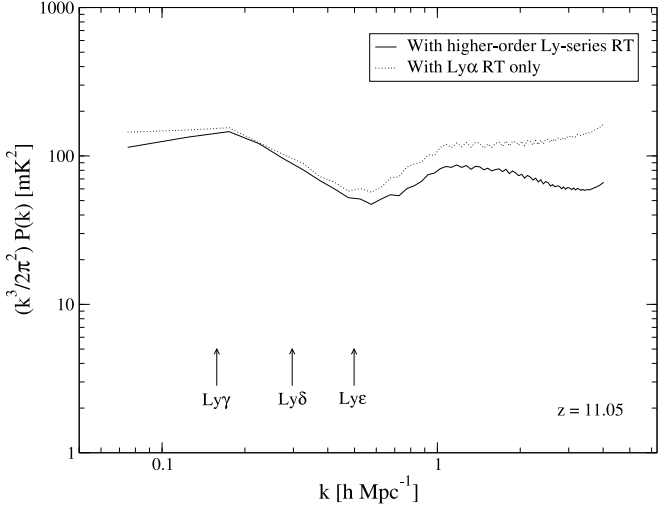}}
\caption{Power spectrum of \dtb at $z = 11.05$ for the fiducial $200 \ h^{-1}$ Mpc simulation (solid line) and for a simulation neglecting higher-order Lyman-series radiative transfer (dotted line). The wavenumbers corresponding to the size of the three horizons are indicated.}
\label{powersp}
\end{figure}

The 3D power spectrum has been widely used as an efficient probe of the \dtb fluctuations. However, we will show that this analysis tool cannot help us in detecting Lyman horizons in our simulated data. In Fig. \ref{powersp}, the \dtb 3D power spectrum of our fiducial $200 \ h^{-1}$ Mpc simulation (solid line) is compared, at $z = 11.05$, with the power spectrum of another simulation that does not take into account higher-order Lyman-series radiative transfer for the computation of \dtb (dotted line). More precisely, we propagate in the latter run the same number of photons, but this time in the interval $[\nu_\alpha,\nu_\beta]$. Arrows indicate the wavenumbers corresponding to the predicted sizes of the Ly$\gamma$, \Lyd and \Lye horizons, as given by equation (\ref{hor2}). The redshift $z = 11.05$ has been chosen because, as we will demonstrate later, this is when other analysis tools provide the best detection of the Lyman horizons. Also, the H\textsc{ii} bubbles at that moment in the simulations are numerous enough to have reliable statistics. As a consequence, the 3D power spectrum is expected to have a trustworthy shape and to be little affected by boxsize effects. On scales greater than the horizons, both spectra are very close to each other. The gain in power, compared with the fiducial simulation, on scales shorter than the Lyman horizons, is caused by the combination of (i) the fact that, compared with the fiducial simulation, we do not loose any photons in the radiative cascades, and (ii) the $r^{-2.3}$ profile of $P_\alpha$ on small scales due to \Lya wing scatterings \citep{Semelinetal07,ChuzhoyZheng07}. Note that the power spectrum shown in this work does not look like theoretical predictions \citep[see e.g.][]{NaozBarkana08}. We conclude from Fig. \ref{powersp} that the two spectra have almost exactly the same shape, illustrating clearly the inability of 3D power spectra to help us in detecting Lyman horizons.

\subsection{Lyman horizons around individual sources}
\begin{figure}[!t]
\resizebox{\hsize}{!}{\includegraphics[angle=270]{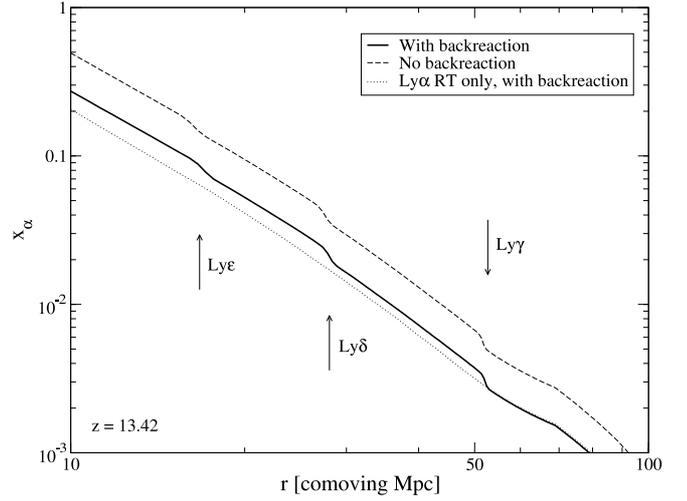}}
\caption{Spherically averaged profile of the coupling coefficient for \Lya pumping $x_\alpha$ at $z = 13.42$, around the first light source in the $100 \ h^{-1}$ Mpc simulation. The correct profile is represented by the thick solid line, while the other two neglect backreaction (dashed line) or radiative transfer for the higher-order Lyman-series lines (dotted line). As expected we observe steep decreases in the value of the profile at radii corresponding to the predicted Ly$\gamma$, \Lyd and \Lye horizons (shown by arrows).}
\label{xa}
\end{figure}

Figure \ref{xa} shows the spherically averaged $x_\alpha$ profile at $z = 13.42$, around the first source of the $100 \ h^{-1}$ Mpc simulation, appearing at $z = 14.06$. The dotted line is the profile obtained when calculating radiative transfer %for photons in the same frequency range $[\nu_\alpha,\nu_\zeta]$, but 
including the \Lya resonance only. In other words, this neglects the contribution of radiative cascades to the total \Lya flux. The dashed line is the result of the correct radiative transfer calculation, including the five Lyman resonances from \Lya to Ly$\epsilon$, but without considering the effect of backreaction. Finally, the solid thick line is the correct averaged $x_\alpha$ profile, with both \Lyn radiative transfer and backreaction. The reason why there is no \Lyb horizon is related to the selection rules forbidding \Lyb photons to convert to \Lya photons. We clearly observe discontinuities in the $x_\alpha$ coefficient profile at the predicted positions.

\begin{figure}[!t]
\resizebox{\hsize}{!}{\includegraphics[angle=270]{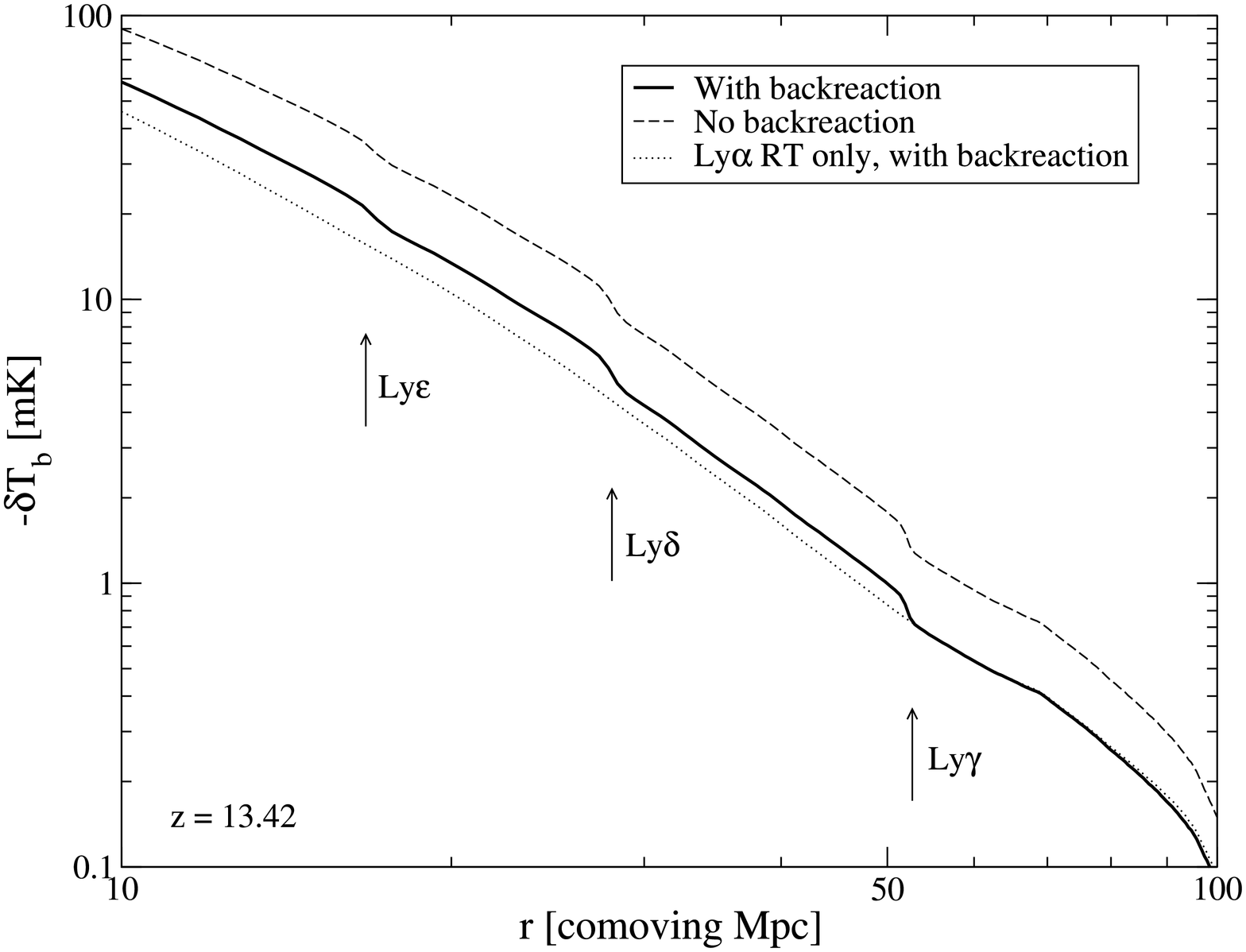}}
\resizebox{\hsize}{!}{\includegraphics[angle=270]{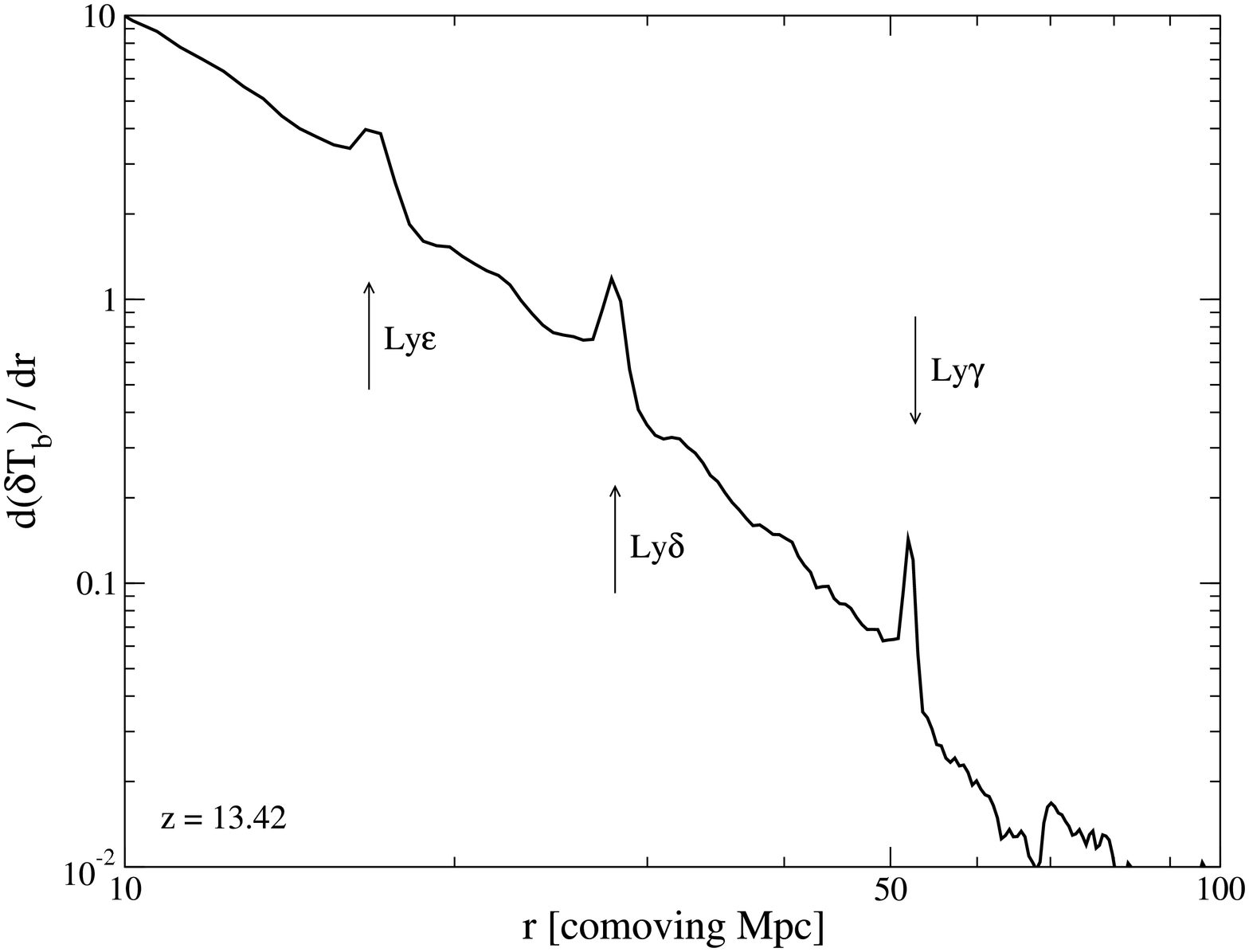}}
\caption{\textit{Upper panel:} Spherically averaged profile of the differential brightness temperature at $z = 13.42$, around the first light source in the $100 \ h^{-1}$ Mpc simulation. The same notation is used as in Fig. \ref{xa}. Note that we plot the opposite value of \dtb. \textit{Lower panel:} Gradient of the spherically averaged profile.}
\label{dtb}
\end{figure}

The corresponding profile for the differential brightness temperature is given in the upper panel of Fig. \ref{dtb}, using the same notation as in Fig. \ref{xa}. Inside the tiny H\textsc{ii} bubble, i.e. in the first few comoving Mpc around the source (not shown in the figure), the signal is in emission. But outside this very small ionized region, the gas kinetic temperature is clearly lower than the CMB temperature at this early redshift. For that reason, the 21 cm signal appears in absorption. It reaches a minimal value of $\sim - 190$ mK at 4.3 comoving Mpc and is as large as about -60 mK at 10 comoving Mpc from the source. Inclusion of higher-order Lyman-series radiative transfer leads to a significant increase by 25\% of $\left(- \delta T_{\mathrm{b}} \right)$ at 10 comoving Mpc compared to the case where only \Lya radiative transfer is considered. Because the \WF effect is dominant  for determining the spin temperature, the discontinuities in the \Lya flux translate into similar discontinuities in the differential brightness temperature profile. At this redshift, $\left(- \delta T_{\mathrm{b}} \right)$ decreases by about 5 mK, 2 mK and 0.2 mK at the Ly$\epsilon$, \Lyd and \Lyg horizon locations respectively. This plot also shows the importance of taking into account backreaction for the correct computation of the spin temperature. Indeed, the box-average kinetic temperature is $\left< T_{\mathrm{k}} \right> = 4.63$ K at $z = 13.42$, and at such a low temperature the backreaction correction factor is about 0.6. In the lower panel of Fig. \ref{dtb}, we plot the gradient of the spherically averaged profile. Steep decreases in the temperature profile at the positions of the Lyman horizons result in prominent peaks in its gradient that make these horizons even easier to detect.
The most important conclusion from Fig. \ref{dtb} is that the different steps in the $x_\alpha$ profile at the Lyman horizons correspond to similar, clearly seen decreases in the differential brightness temperature profile.

\begin{figure}[!t]
\resizebox{\hsize}{!}{\includegraphics{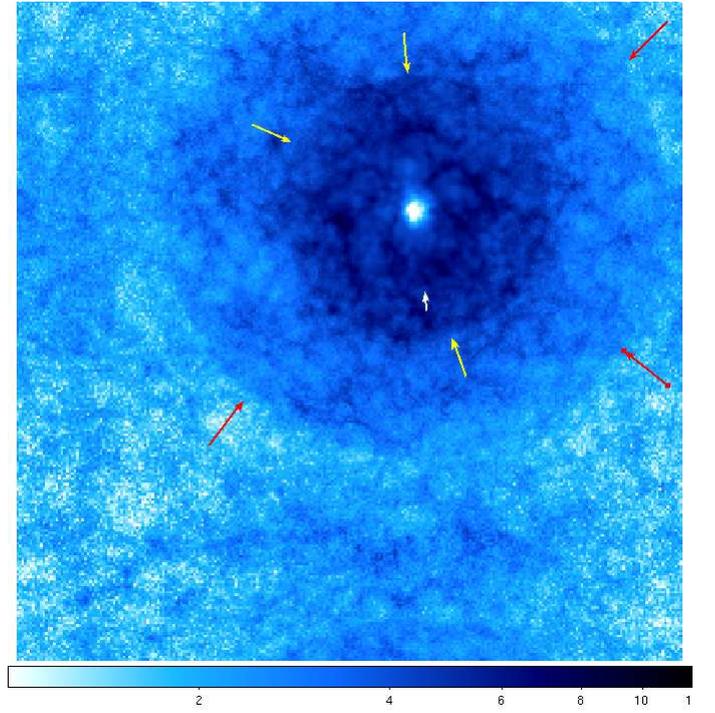}}
\caption{Map of the quantity $- \delta T_{\mathrm{b}} \times r^2$ at $z = 13.42$, with $r$ the distance to the source center, in arbitrary units. The Ly$\epsilon$, \Lyd and \Lyg horizons are marked with white, yellow and red arrows respectively. The size of the box is $100 \ h^{-1}$ Mpc and the slice thickness is $2 \ h^{-1}$ Mpc. The color scale is logarithmic.}
\label{map}
\end{figure}

We show in Fig. \ref{map} a map of the quantity $- \delta T_{\mathrm{b}} \times r^2$ around the same source at the same redshift $z = 13.42$, with $r$ the distance to the source center. Mapping this quantity enhances visualization. Indeed, from equations (\ref{equts}) and (\ref{equdtb}), and since\footnote{More precisely, taking wing scatterings into account slightly steepens the profile on small scales, as noted above \citep{Semelinetal07,ChuzhoyZheng07}.} $x_{\alpha} \propto r^{-2}$, it can easily be shown that $\delta T_{\mathrm{b}}$ is also proportional to $r^{-2}$. Mapping the product $- \delta T_{\mathrm{b}} \times r^2$ will thus straighten up the radial profiles, making the horizons more apparent. The slice thickness is $2 \ h^{-1}$ Mpc and the map scale is logarithmic. The three spherical, concentric horizons are marked with white, yellow and red arrows for the Ly$\epsilon$, \Lyd and \Lyg discontinuities respectively. On this map, the \Lyd horizon is particularly obvious. Note the perfectly spherical shape that characterizes these features.

\subsection{Effect of the nearby sources}
\begin{figure*}[!t]
\centering
\includegraphics[angle=270,width=17cm]{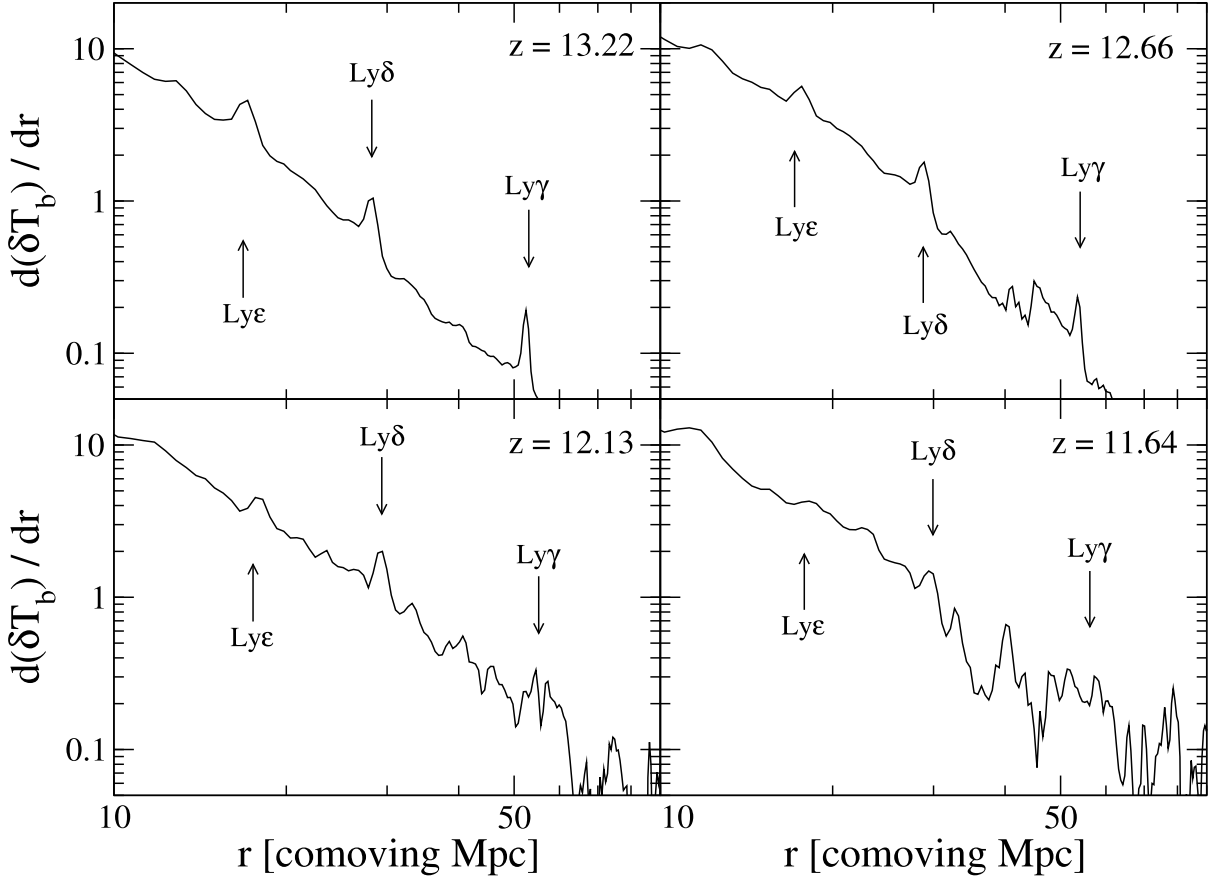}
\caption{Gradient of the spherically averaged profiles of the differential brightness temperature, around the first light source appearing at $z = 14.06$ in the $100 \ h^{-1}$ Mpc simulation. The four panels show the profile at different redshifts, from $z = 13.22$ to $z = 11.64$. At each redshift, arrows show the predicted position of the Ly$\epsilon$, \Lyd and \Lyg horizons. The \Lye and \Lyd horizons can be detected during a redshift interval $\Delta z \sim 2$. The fainter \Lyg horizon is visible for a shorter period $\Delta z < 1.5$.}
\label{prof_dtb_multi}
\end{figure*}

From Figs. \ref{xa} to \ref{map}, we are able to conclude that the existence of Lyman horizons around the first light sources create similar features in the 21 cm signal at the very beginning of the EoR. However, these weak discontinuities, whose amplitude is only a few mK, will be progressively affected by the contribution of other nearby sources located closer than the $\sim 50$ comoving Mpc corresponding to the farthest \Lyg horizon. Also, Lyman photons emitted between \Lya and \Lyg by more distant sources can travel more than 50 comoving Mpc, and thus are likely to reach the outskirts of many other sources.
%This is illustrated in Fig. \ref{xa_43}, where we show the spherically averaged $x_{\alpha}$ profile around the first source at redshift $z = 12.30$. At that time, eight other sources have appeared, resulting in eight peaks in the profile at the corresponding distances to the first source.
The increasing influence of the other UV emitters on the brightness temperature profile of an individual source is shown in Fig. \ref{prof_dtb_multi} for the $100 \ h^{-1}$ Mpc simulation box, from $z = 13.22$ to $z = 11.64$. This redshift interval is covered by 10 snapshots and the number of emitting cells is 6 at $z = 13.22$, 15 at $z = 12.66$, 38 at $z = 12.13$, and 86 at $z = 11.64$. Early in the simulation, the three Lyman horizons are clearly seen as prominent peaks in the gradient of the temperature profile (upper left panel). Later, as the influence of the other sources becomes stronger, fluctuations in the profile grow until they reach the same amplitude as the horizons. At the beginning of this process, some of the horizon peaks are still observable (upper right and lower left panels), but after a certain time they are not distinguishable anymore (lower right panel). We find that for this particular simulation the \Lye and \Lyd horizons are observable during a redshift interval $\Delta z \sim 2$ and that the size of these discontinuities is about 2-4 mK. The \Lyg horizon is visible for a shorter period $\Delta z < 1.5$.
%\begin{figure}[!t]
%\resizebox{\hsize}{!}{\includegraphics[angle=270]{profile_xa_hir_THERM_100_JADE_analyse_43.eps}}
%\caption{Spherically averaged profile of the coupling coefficient for \Lya pumping $x_\alpha$ around the first source at redshift $z = 12.30$. The strong influence of the other eight sources of the simulation box present at that time is visible as sharp peaks in the profile.}
%\label{xa_43}
%\end{figure}

The size of our simulation box can also be important in this attenuation process. Because of the periodic boundary conditions, a photon emitted, say, between \Lya and \Lyb can travel a distance greater than the box size, resulting in a self-contamination of the \Lya flux. Also, the smaller the simulation box, the more homogeneous is the distribution of sources around any given point in the box. In order to reduce these effects, we considered the larger, $200 \ h^{-1}$ Mpc simulation and find that, despite the larger volume, the Lyman horizons around the first source of the simulation are observable during about the same redshift range as for the first source in the $100 \ h^{-1}$ Mpc simulation: $\Delta z \sim 2$ for the \Lye and \Lyd discontinuities, $\Delta z \sim 1.5$ for the \Lyg step. These are maximal intervals and sources lighting up later will have their horizons observable for a reduced period. As an example, the seventh source in the $200 \ h^{-1}$ Mpc simulation, appearing at $z=13.03$, has $\Delta z \approx 1.2 - 1.5$ for \Lye and Ly$\delta$, and the \Lyg horizon can be detected in the first few snapshots only.

The \Lye and \Lyd horizons are thus the best candidates for detection, the latter one being visible a little bit longer. The \Lyg discontinuity is weaker because of the $r^{-2}$ scaling of the \Lya flux, and is thus more sensitive to the influence of the other sources.
%On the other hand, the difficulty to observe the \Lye horizon could be the result of the highly non-symetric density field near the source.
%Because we use the same total number of photons per snapshot in both the $100 \ h^{-1}$ Mpc and the $200 \ h^{-1}$ Mpc simulations, the number of photons per source is smaller in the larger box. For that reason, comparison of the redshift interval during which Lyman horizons are visible is not straightforward. But it seems that a period $\Delta z \sim 2$ around $z = 13$ is a maximal value. As soon as contamination by the other sources becomes important, this interval decreases.

The growing fluctuations appearing in the spherically averaged temperature profiles clearly originate from the \Lya background that develops as more and more sources light up. At first sight, density or gas temperature fluctuations could also be suspected (see equation \ref{equdtb}). We thus ran test simulations in which we computed \dtb assuming a perfectly homogeneous and isothermal IGM in the Lyman radiative transfer post-processing phase. The density was chosen to be the critical density and the gas kinetic temperature was assumed to be equal to the box-averaged value of the normal run. In this way, only the \Lya flux was likely to let \dtb vary from one point to the next. The results of this test showed that the Lyman discontinuities are not seen for a significantly longer period than in the standard case. Indeed, we were able to identify the three horizons in a few more snapshots only, and we conclude that the density and temperature anisotropies around the source are not responsible for the disappearance of the Lyman features. A final possibility in explaining the vanishing of the Lyman steps is Monte-Carlo noise. Indeed, the total number of photons per snapshot is constant, but the source number grows quickly. As a result, the number of photons per source decreases very rapidly. However, using very expensive simulations with ten times more photons between each pair of snapshots ($16 \times 10^9$) in both the $100 \ h^{-1}$ and $200 \ h^{-1}$ Mpc simulations results in undistinguishable radial profiles. We thus clearly demonstrated that the Lyman horizons disappear around individual sources because of the growing influence of the \Lya flux of the nearby sources.

\subsection{Stacking}
\begin{figure}[!t]
\resizebox{\hsize}{!}{\includegraphics[angle=270]{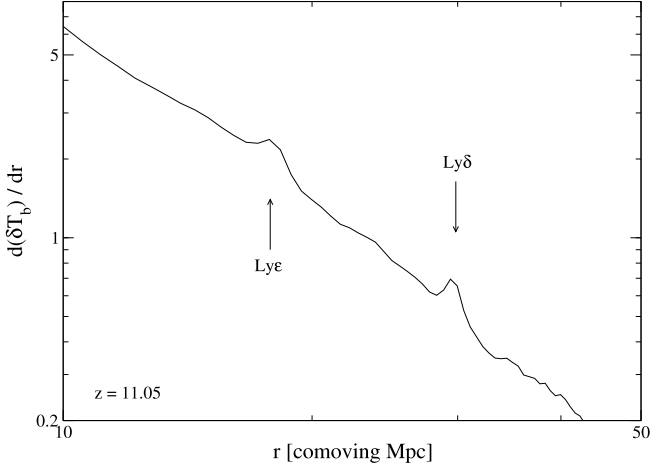}}
\caption{Gradient of the stacked \dtb radial profile at $z = 11.05$ for the $200 \ h^{-1}$ Mpc simulation. The stacked profile is obtained by averaging the individual profiles of all the source-containing cells at this redshift. This method allows us to detect the \Lye and \Lyd horizons, whose predicted positions are indicated by arrows.}
\label{grad_stack}
\end{figure}

In order to extend the detectability of the Lyman steps to lower redshifts, we considered a stacking technique. This allows one to smooth the fluctuations present in the individual profiles and to strengthen the visibility of the horizon steps. As an example to illustrate the efficiency of this method, we stacked the \dtb radial profiles around all the cells that contain sources in the interpolation grid of the $200 \ h^{-1}$ Mpc simulation, at $z = 11.05$, when the Lyman horizons are undetectable around the vast majority of the individual sources. For some sources, small features are seen at the predicted horizon positions, but are indiscernible from fluctuations of similar amplitude. At this redshift we find 1433 source-containing cells. The individual profiles are simply added and the result is divided by the number of cells we are considering\footnote{We can safely ignore the fact that the stacked sources have different peculiar velocities perpendicular to the line of sight. Indeed, a rapid estimation shows that a peculiar velocity about 300 km s$^{-1}$ leads to a redshift difference $\Delta z = (1+z) \ v/c \sim 10^{-2}$. As shown in Fig. \ref{fighor} and equation (\ref{hor2}), this results in completely negligible differences in the horizon sizes.}. Figure \ref{grad_stack} shows the gradient of the stacked profile. The presence of the \Lye and \Lyd horizons is now obvious.
We find that this method allows us to detect these two discontinuities unambiguously down to redshifts close to 10. Interestingly however, the stacking method does not provide good results for the smaller simulation. We believe that this comes from the number of source-containing cells being about five times smaller in the $100 \ h^{-1}$ Mpc run at a given redshift. This has the consequence that the number of profiles that are averaged is not enough to smooth the individual fluctuations.

\subsection{Detectability}
We would now like to examine the possibility for detecting the Lyman horizons with the planned Square Kilometre Array. To reach this goal we have to include in our simulated data both the instrument resolution and noise. We choose the noise power spectrum given by equation (12) of \citet{Santosetal11}, which models the resulting noise after foreground removal has been performed:
\begin{equation}
P_N(k,\theta) = r^2 y \frac{\pi \lambda^2 D^2_{\rm max} T^2_{\rm sys}}{t_0 A^2_{\rm tot}},
\end{equation}
where $k$ is the wavenumber, $\theta$ the angle between the wave vector and the line of sight, $r(z)$ the comoving distance to redshift $z$, $y$ a conversion factor between frequency intervals and comoving distances, $\lambda$ the observation wavelength, $D_{\rm max}$ the maximum baseline, $T_{\rm sys}$ the system temperature, $t_0$ the total observation time, and $A_{\rm tot}$ the total collecting area. Following Santos et al., we use $D_{\mathrm{max}} = 10$ km, $t_0 = 1000$ hours and a system temperature modelled by their equation (13) as the sum of the receiver noise temperature (assumed to be 50 K) and the sky temperature (dominated by the Galactic synchrotron emission):
\begin{equation}
T_{\rm sys}(z) = 50 + 60 \left( \frac{1 + z}{4.73} \right)^{2.55} {\rm K}.
\end{equation}
We compute the total collecting area using the sensitivity value of 4000 m$^2$/K, in agreement with the design reference for the instrument. Following these authors, we also use 70\% of $A_{\rm tot}$ in the noise calculation, considering that the rest of the collecting area is used for point source removal and calibration.

\begin{figure}[!t]
\resizebox{\hsize}{!}{\includegraphics[angle=270]{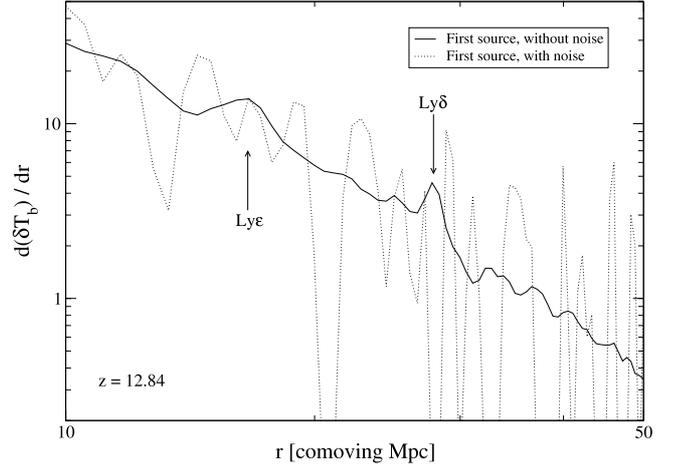}}
\caption{Gradient of the \dtb profile around the first source appearing in the $200 \ h^{-1}$ Mpc simulation at $z = 12.84$, with (dotted line) and without (solid line) SKA-like noise.}
\label{ancienne_fig}
\end{figure}

Figure \ref{ancienne_fig} illustrates the dramatic effect that noise may have on the detectability of Lyman horizons. The plotted example shows the gradient of the \dtb radial profile around the first source of the $200 \ h^{-1}$ Mpc simulation at $z = 12.84$, with (dotted line) and without (solid line) noise. While the \Lye and \Lyd horizons are clearly detected in the noise free profile, this is no longer true when noise is added.

However, the field of view (FoV) subtended by our simulation (less than 3 deg$^2$ in the range $z = 11 - 14$) is much smaller than probable SKA field of view. If the ratio between both fields is $N$, then noise will be reduced in the stacking procedure by a factor $\sqrt{N}$ larger for the SKA FoV than for the simulation FoV. We will here consider a field of view of 400 deg$^2$ for the SKA. To simulate the efficiency of the stacking procedure for such a large field of view, we decrease the noise level in the simulation FoV by the appropriate $\sqrt{N}$ factor. The fact that we stack over the \textit{volume} of our simulation cube is not problematic in this analysis. Indeed, a short calculation shows that the full length of our $200 \ h^{-1}$ Mpc simulation represents a redshift difference $\Delta z \sim 1.5$, which corresponds to a $\sim 6 \%$ difference in the horizon sizes at $z \sim 13$, according to equation (\ref{hor2}). For example, this represents $1.6$ comoving Mpc for Ly$\delta$, or about $0.5\arcmin$.

\begin{figure*}[!t]
\centering
\includegraphics[width=17cm]{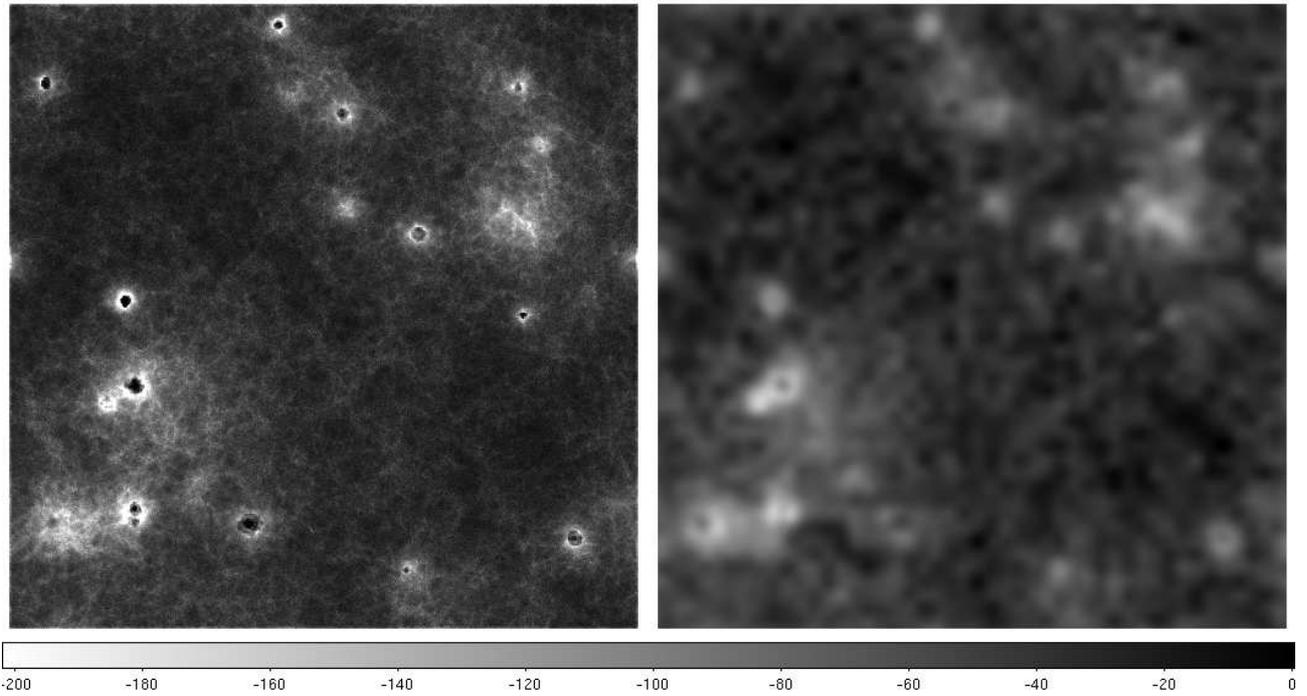}
\caption{Maps of \dtb at $z = 11.05$. The slice is $200 \ h^{-1}$ Mpc on a side and has a thickness of $2 \ h^{-1}$ Mpc. The scale, in units of mK, is linear. The left panel shows our simulated data cube at full resolution. The right panel shows the effect of instrumental noise and instrumental resolution. At that redshift, the latter is slightly larger than $2\arcmin$.}
\label{map_noise_res}
\end{figure*}

The most limiting factor will probably be the resolution of the instrument. The proposed SKADS-SKA implementation for observations between 70 MHz and 450 MHz suggests a 5km-diameter core\footnote{\texttt{http://www.skads-eu.org/PDF/SKADS\_White\_Paper\_100318\\\_dist.pdf}}. This gives a resolution larger than $2\arcmin$ in the redshift range $z = 11-14$ relevant for our study. In order to mimic the effect of the limited instrumental resolution, we apply 3D\footnote{For simplicity reason, we consider the same resolution in the frequency direction as in the other two spatial directions.} gaussian smoothing with a FWHM equal to the resolution. We compare in Fig. \ref{map_noise_res} a map at $z = 11.05$ of the full-resolution, noise-free simulated data (left panel) with the same field when one includes noise and convolution with the appropriate resolution (right panel). The full-resolution map shows a characteristic signature of the early EoR 21 cm signal in the vicinity of the sources, with practically spherical H\textsc{ii} bubbles being surrounded by absorption regions with deep \dtb local minima. On the other hand, only some of the largest bubbles are still detected in the convolved map. This will have an influence on the identification and localization of the first sources, and hence on attempts to observe Lyman horizons. Also, the limited resolution of the future observations will cause a blurring of the small discontinuities we want to detect.

\begin{figure}[!t]
\resizebox{\hsize}{!}{\includegraphics[angle=270]{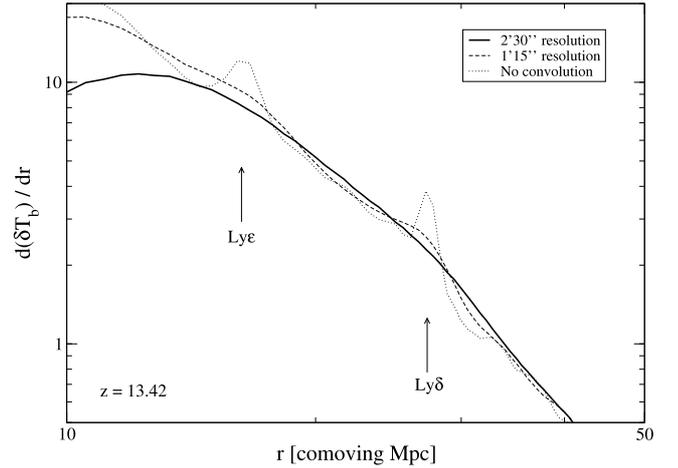}}
\caption{The thick solid (thin dashed) line shows the gradient of the \dtb stacked profile in the $200 \ h^{-1}$ Mpc simulation at $z = 13.42$, with SKA-like noise, assuming a field of view of 400 deg$^2$, and a $2\arcmin30\arcsec$ ($1\arcmin15\arcsec$) resolution. For comparison, we also show the gradient of the \dtb stacked profile, at the same redshift, without noise nor resolution effect (dotted line).}
\label{res_noise}
\end{figure}

Finding an algorithm to efficiently detect the H\textsc{ii} bubbles in the convolved maps will be the subject of a future paper. Here, we tackle the question of pinpointing the source positions on the sky beyond the resolution accuracy, which could leave its mark on the stacking procedure, in the following way. Instead of using the exact source position when building the stacked profile, we randomly distribute its location inside a sphere centered on the exact position. The diametre of that sphere corresponds to the SKA resolution at the considered redshift. Figure \ref{res_noise} shows the gradient of the stacked profile assuming the observation strategy described above (thick solid line), compared to the case without noise and with full resolution, and in which the position of the sources is perfectly known (dotted line). The important field of view allows us to significantly reduce the effect of the noise, but the $2\arcmin$ resolution has a dramatic effect on the Lyman horizons, which are completely wiped off. The dashed line shows that with a $1\arcmin$ resolution (10 km core), the \Lyd horizon is detected as a wide and weak hump. In order to maximize the detection probability, we assumed here perfect foreground removal and chose a redshift ($z = 13.42$) at which the Lyman horizons are not yet contaminated by the other nearby sources. We deduce from this figure that detection of the Lyman horizons in the early phases of the EoR using SKA observations will probably be possible only with a $1\arcmin$ resolution. 

%The analysis we proposed here was simplified by our a priori knowledge of the exact positions, at any time, of all the sources in our simulations. In reality, construction of the stacked profiles from future radio interferometric observations will be more challenging, because we will first have to precisely determine these positions.  Thus, there are great hopes of finding an efficient algorithm in order to determine the exact position of the sources. This will be the subject of a future paper.
%---------------------------------------------------------------------------------------

%---------------------------------------------------------------------------------------
\section{Discussion and conclusions}
\label{concl}
%---------------------------------------------------------------------------------------
In this paper we studied the effects of higher-order Lyman-series radiative transfer on the differential brightness temperature of the 21 cm signal of neutral hydrogen during the early stage of the EoR. This signal was computed in the presence of inhomogeneous Wouthuysen-Field effect using the Monte-Carlo 3D radiative transfer code LICORICE. We showed that the discontinuities in the Ly$\alpha$ flux radial profile of the first light sources, caused by the existence of discrete Lyman horizons, lead to similar features in the differential brightness temperature profile.

We found that the \Lye and \Lyd horizons are detected in our simulations around individual sources during a redshift interval $\Delta z \sim 2$ after the first source lights up. Later, the \Lya flux at a given point has a growing smooth background component due to the cumulative contribution of the other sources, and the Lyman steps fade away. However, we showed that stacking the individual profiles makes the visibility period significantly longer ($\Delta z \sim 4$). 

%Two important caveats have to be mentioned here.The first one comes from the mass resolution of our simulations, in which the first sources appear quite late, at $z \sim 14$. A better resolution would imply earlier source formation, and hence would change the corresponding $(1 + z )$ 21 cm observed frequency. If the first sources appear too early, there is a danger that the associated observations would be out of the frequency range of the future instruments. The stacking method used in our analysis could play here a positive role in extending the frequency range of detectability.

An important caveat has to be mentioned here. We have considered in our simulations purely stellar sources. However, it should be kept in mind that considering an important contribution from quasars would result in the \WF effect being significantly induced by \Lya photons coming from collisional excitation of hydrogen atoms by secondary electrons produced by X-rays. If an important part of the total number of \Lya photons is not coming from redshifting and cascading photons, the interesting features in \dtb we discussed in this work would be more difficult to extract.

On the other hand, let us note that the size of the Lyman discontinuities could increase if the number of \Lya photons from radiative cascades is enhanced. In a recent paper, \citet{Meiksin10} claims that the number of such photons could be up to 30\% higher than previously considered. In this case, the size of the steps in the $x_{\alpha}$ profile would increase interestingly, together with the size of the corresponding \dtb discontinuities.

Detections of these discontinuities would be of first importance for two reasons. First, such features would be very interesting for researchers involved in foreground removal. Indeed, removing the much brighter foregrounds (Galactic synchrotron and free-free emission, extragalactic sources, ionospheric distortions) will be a serious issue for future 21 cm observations. But Lyman horizons are structures whose size is precisely known at a given redshift (or frequency), and for that reason positive detections would ensure that the laborious removal task has been correctly handled. The second reason is the possible use of these horizons as standard rulers, as was noted by \citet{PritchardFurlanetto06}.

%Considering instrumental noise results in suppressing the horizons around individual sources at the highest redshifts, but we showed that as soon as the number of sources in the simulation is sufficient, the stacked profiles are barely affected by the noise component. Since observations will cover a much larger region than our simulations (the $200 \ h^{-1}$ Mpc box is $1.5\degr \times 1.5\degr$ at $z \approx 10$), stacking may be efficient even at the very beginning of the EoR.

When assessing the question of detecting Lyman horizons with the planned Square Kilometre Array by taking into account the reference implementation and design, we found that such a detection will be very challenging. Indeed, the horizons are not detected with a $2\arcmin$ resolution. A $1\arcmin$ resolution may allow us to detect the \Lyd horizon.
%---------------------------------------------------------------------------------------

%---------------------------------------------------------------------------------------
\begin{acknowledgements}
We are grateful to J. Uson for helpful discussion. We also thank the anonymous referee for his/her insightful comments. This work was realized in the context of the LIDAU project. The LIDAU project is financed by a French ANR (Agence Nationale de la Recherche) funding (ANR-09-BLAN-0030). PV acknowledges support from a Swiss National Science Foundation (SNSF) post-doctoral fellowship. This work was performed using HPC resources from GENCI-[CINES/IDRIS] (Grant 2011-[x2011046667]).
\end{acknowledgements}
%---------------------------------------------------------------------------------------

%---------------------------------------------------------------------------------------
\bibliographystyle{aa} % style aa.bst
\bibliography{lyalphabib}
%---------------------------------------------------------------------------------------

\end{document}